\documentclass[11pt,a4paper,twocolumn]{article}

\usepackage{graphics,amsmath,graphicx,amssymb}
\usepackage{times}
\usepackage[margin=1.5cm]{geometry}

\newenvironment{sciabstract}{%
\begin{quote} \bf}
{\end{quote}}

\title{The origin of Mooij correlations in disordered metals}

\author{Sergio Ciuchi$^{1,2}$, Domenico Di Sante$^{3}$, Vladimir Dobrosavljevi\'c$^4$ \\ \&
 Simone Fratini$^{5\ast}$\\
\\
\normalsize{$^1$Department of Physical and Chemical Sciences, University of L'Aquila, }\\ \normalsize{ Via Vetoio, L'Aquila, Italy I-67100}\\
\normalsize{$^2$Consiglio Nazionale delle Ricerche (CNR-ISC)  Via dei Taurini, Rome, Italy I-00185}\\
\normalsize{$^3$Institute of Physics and Astrophysics, University of W\"urzburg,   W\"urzburg, Germany}\\
\normalsize{$^4$Department of Physics and National High Magnetic Field Laboratory, }\\ \normalsize{  Florida State University, Tallahassee, Florida 32306, USA}\\
\normalsize{$^5$Institut N\'{e}el-CNRS and Universit\'{e}
 Grenoble Alpes, }\\ \normalsize{  Bo\^{i}te Postale 166, F-38042 Grenoble Cedex 9, France}\\
 \\
\normalsize{$^\ast$To whom correspondence should be addressed; E-mail:  simone.fratini@neel.cnrs.fr.}}
\date{}

\begin{document} 



\twocolumn[
  \begin{@twocolumnfalse}
\maketitle


\begin{sciabstract}
Sufficiently disordered metals display systematic deviations from the behavior
predicted by semi-classical Boltzmann transport theory. Here the scattering events from
impurities or thermal excitations can no longer be considered as additive
independent processes, as asserted by Matthiessen's rule following from this picture. 
In the intermediate region between the regime of good conduction and that of
insulation, one typically finds a change of sign of the temperature
coefficient of resistivity (TCR), even at elevated temperature spanning ambient
conditions, a  phenomenology that was first identified by Mooij in 1973.
Traditional weak coupling approaches to identify relevant corrections to the
Boltzmann picture focused on long distance interference effects such as
"weak localization", which are especially important in low dimensions (1D, 2D) 
and close to the zero temperature limit.
Here we formulate a strong-coupling approach to tackle the
interplay of strong disorder and lattice deformations (phonons) in bulk three-dimensional metals
at high temperatures. We identify a
polaronic mechanism of strong disorder renormalization, which describes how a
lattice locally responds to the relevant impurity potential. This mechanism,
which quantitatively captures the Mooij regime, is physically distinct and
unrelated to Anderson localization, but realizes early seminal ideas of Anderson
himself, concerning the interplay of disorder and lattice deformations. 

\end{sciabstract}

\end{@twocolumnfalse}]

\section*{Introduction}

Identifying physical mechanisms that control electronic transport in materials 
has long been recognized as 
a central task in condensed matter physics. In good conductors, for example, it is known that
finite electrical resistivity arises due to scattering processes from impurities
or various thermal excitations, as very successfully described by the
Boltzmann theory of transport. Here, the various scattering events can be
considered as statistically independent and thus additive, leading to the
well-known Matthiessen's rule, where any thermally-induced scattering simply
{\em increases} the resistivity $\rho (T)$. This behavior, which is well
documented in most metals, corresponds to a positive Temperature Coefficient
of Resistivity (TCR), i.e.  $ d\rho/dT > 0$.

At stronger disorder, where the residual resistivity $\rho_0 =  \rho (T=0)$
increases towards the Mott-Ioffe-Regel (MIR) limit \cite{fisk76prl,Hussey,CalandraPRB02,GunnarrssonCalandraRMP2003}
 (corresponding to short scattering lengths on the atomic scale), significant deviations to
the weak scattering picture are commonly observed, eventually leading to a change of
sign of the TCR \cite{Mooij,LeeRMP}. 
One popular view on the possible cause for this behavior
\cite{Kaveh,Tsuei,Imry,Gantmakher} invokes a
mechanism for impurity-induced bound electronic state formation at strong
disorder --- the so-called Anderson localization \cite{Anderson}. This picture 
involves a modification of the electronic wavefunctions due to interference processes, but it
does not envision any significant rearrangement of the electronic spectra --- no gap needs to open.
The related "weak localization" corrections prove to describe well
the leading low-temperature behavior in metals at weak disorder, where careful
perturbative treatments have provided a consistent theoretical picture. This is the
regime where such interference processes dominate, but it is by no means obvious
that they do so at stronger disorder or around room temperature.

On the other hand, a surprisingly robust phenomenology describing how the TCR
changes sign around the MIR limit has been established by Mooij in the early 1970s \cite{Mooij},
noticing that the slope of the resistivity curves 
linearly (anti)correlates with the extrapolated zero-temperature value $\rho_0$,
which has subsequently been confirmed on hundreds of materials (i.e. essentially every metal 
where a sufficient degree of disorder could be experimentally achieved) \cite{LeeRMP,Tsuei,Nagel,Naugle,Hussey}.
Most remarkably, such apparent universality was found in the high temperature regime,
often extending to hundreds of Kelvin. This observation
is important, since non-local interference processes underpinning (weak)
localization require long-distance phase coherence, a situation that can hardly
be expected to hold at elevated temperatures where incoherent thermal
excitations abound. Indeed, the transport behavior of metals in this temperature
regime is known to be dominated by lattice vibrations (phonons), leading to the
familiar linear-T resistivity above the Debye temperature. The aim of this work is to 
propose, and  
validate against available experiments, 
an alternative scenario that can
more plausibly explain the ubiquitous high temperature anomalies identified by Mooij.

To set the stage for our approach, we recall that alternative physical ideas
have been put forward in early work by Anderson himself, in discussing
the possible role of lattice deformations in the limit of strong disorder \cite{AndersonNat72}. He
argued that the reduced mobility of electrons in poor conductors may allow the
lattice deformations to self-trap electrons through a disorder-assisted polaronic
effect. As in other examples of strong interaction effects, this polaronic mechanism 
should result in rearrangements of the electronic energy levels, leading to gap
formation and substantial transfer of spectral weight away from the Fermi
energy. In recent work \cite{DiSantePRL17} we formulated a microscopic theory able 
to capture both Anderson
localization and such strong-coupling polaronic effects. Importantly, it was shown that 
the latter always acquire a dominant role at strong disorder (i.e., surprisingly, 
polaronic deformations naturally arise even in common metals with weak electron-phonon interactions), 
provided that one releases the customary assumption where metals are 
regarded as infinitely rigid, undeformable objects \cite{Anderson,LeeRMP}.
The main physical point of the present
paper is to emphasize that this mechanism, while completely unrelated to Anderson
localization,  holds the key to understanding the universal high
temperature anomalies in the Mooij regime. We do this by formulating a theory
which by construction disregards non-local interference effects associated with
Anderson localization, but does capture strong-coupling physics associated with
disorder-assisted polaron formation. We show that important local correlations
between the impurity potential and the induced lattice deformations directly
cause the breakdown of  Matthiessen's rule, and in fact allow quantitative
description of Mooij correlations found around the MIR limit.

We intend to address the deviations from Matthiessen's rule observed in sufficiently disordered metals
at high temperatures, and to demonstrate
the origin of the existing relation between the 
temperature coefficient of the resistivity, $d\rho/dT$, and its residual zero-temperature  value $\rho_0$. 
To this end we  apply 
dynamical mean-field theory in the coherent potential approximation (DMFT-CPA) ---
an inherently strong-coupling theory
which takes the viewpoint that the dominant physical processes occur at the local level,
and allows us to include both disorder and the desired interaction effects in a reliable way 
\cite{MillisA15} (see the SM file, Sec. 1). The temperature-dependent resistivity 
is then evaluated using the Kubo-Greenwood formula, i.e. explicitly going beyond the semiclassical 
Boltzmann treatment. We provide next 
an analytical derivation illustrating the physical origin of the phenomenon in full generality, and then proceed 
with the discussion of the numerical DMFT-CPA results. 

\section*{Results}
\subsection*{Breakdown of Matthiessen's rule and Mooij correlations}

Let us consider a disordered metal at $T=0$, which we take as our reference system. 
In the  local picture, 
the solution of the disorder problem is entirely characterized by 
the  local Green's function
(or, equivalently, the local density of states).
This quantity, that we denote by $\hat{G}_{el}^\xi(\omega)$,  
varies from site to site depending on the value 
of the local random potential $\xi$, and is entirely determined
from the  knowledge of the statistical distribution 
$P(\xi)$ of site energies. From its averaged value over the sample, $G_{el}(\omega)=\langle \hat{G}_{el}^\xi(\omega) \rangle$, 
one  determines the self-energy $\Sigma_{el}(\omega)$, which incorporates  
the relevant elastic scattering processes related to the random
environment. 
The residual resistivity $\rho_0$ of the metal at $T=0$
is then readily evaluated from the elastic scattering rate
$\Gamma_{el}=-2 Im \Sigma_{el}(\omega=E_F)$, with $\rho_0\propto \Gamma_{el}$ from the Drude theory of metals.

\begin{figure*}[t!]
\centering
\includegraphics[width=14cm]{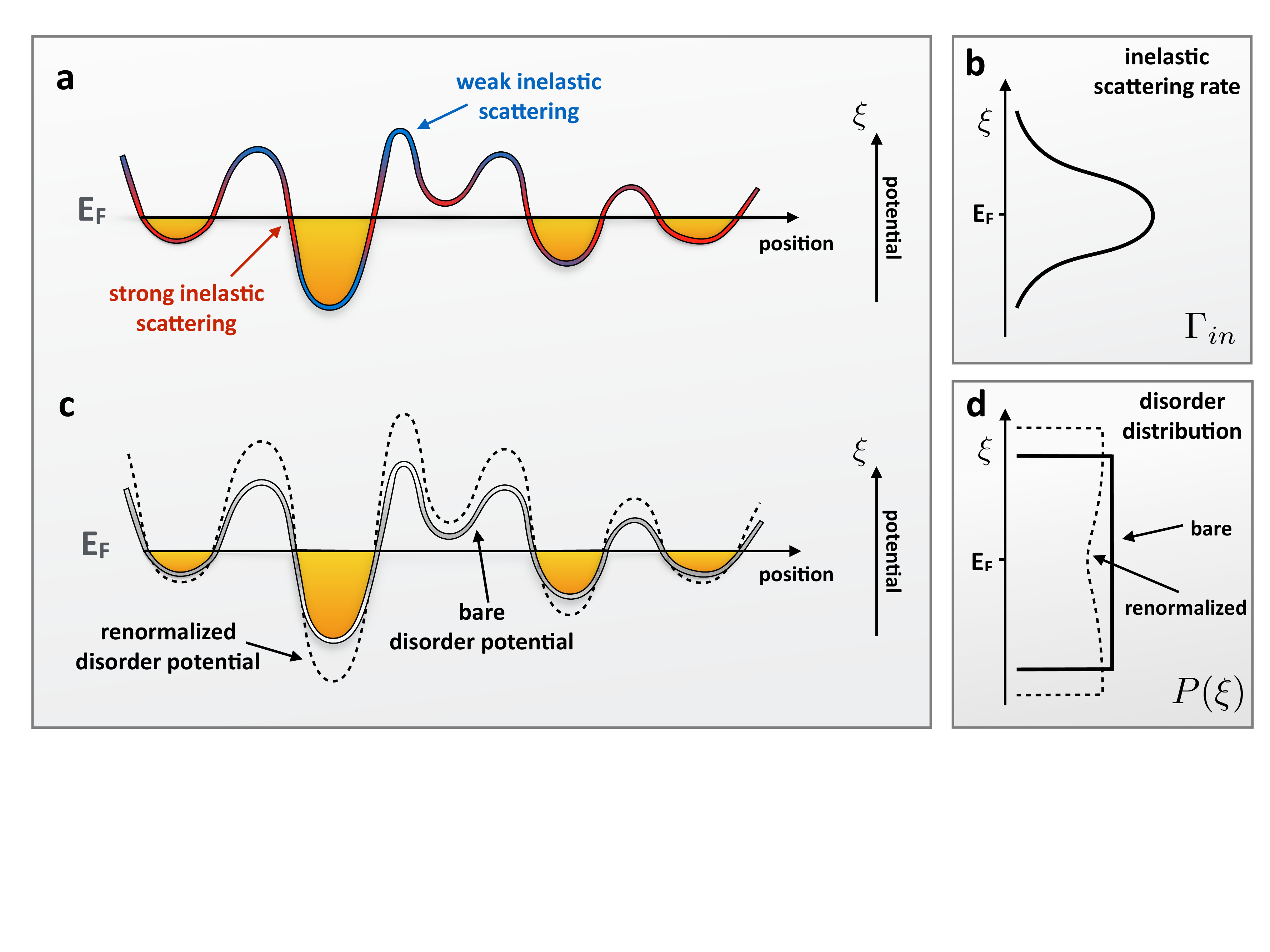}
\caption{\textbf{Breakdown of Matthiessen rule and polaronic renormalization of disorder.}
(a,b) The rate of inelastic scattering correlates with the value of the local 
random potential $\xi$, being maximum 
at the Fermi energy $E_F$ and minimum away from it, as indicated by the color code 
(red: strong scattering, blue: weak scattering). 
(c,d) The deformable lattice responds to the spatial fluctuations of the electron density (shaded yellow), renormalizing the 
random potential (dashed line), which opens a dip in the distribution $P(\xi)$.
}
\label{fig:sketches}
\end{figure*}

The  temperature coefficient of the resistivity can now be addressed by 
performing an expansion in 
the lattice fluctuations, as these are responsible for the leading $\rho \propto T$ term 
(thermal smearing of the Fermi surface is negligible at this stage, as it 
brings corrections $\propto T^2$ and therefore does not affect the TCR, see Fig. (S2)
in the SM file).
Using the Dyson equation we obtain the total 
self-energy, which incorporates both elastic and inelastic scattering, as:
\begin{equation}
\Sigma=\Sigma_{el}+G_{el}^{-1}\left\langle \hat{G}_{el}^\xi \ \hat{\Sigma}_{in}^\xi \ \hat{G}_{el}^\xi
\right\rangle G_{el}^{-1}.
\label{eq:Dyson}
\end{equation}
 (all quantities 
here and in what follows are functions of frequency $\omega$, which we omit for clarity).
The term $\hat{\Sigma}_{in}^\xi$ describes the 
inelastic emission and absorption of
phonons.
 Its explicit dependence 
on the atomic site energy $\xi$ indicates
that the way an electron is affected by such inelastic
 processes is  different from site to site, depending on its 
local random environment, as sketched in Fig. \ref{fig:sketches}(a,b).
Eq. (\ref{eq:Dyson}) reveals how  correlations between disorder and electron-phonon scattering emerge:
 the propagation in the disordered lattice and the interaction processes are
intertwined because they take place in the same 
region of the experimental sample,
symbolized here by the same given value of the site-disorder variable $\xi$.
The formal separation between  scattering channels underlying Matthiessen's rule would arise
only if we were to treat the different terms in Eq. (\ref{eq:Dyson}) as independent processes,
which corresponds to factorizing the averages as $\langle \hat{G}_{el}^\xi\rangle \langle   \hat{\Sigma}_{in}^\xi \rangle 
\langle\hat{G}_{el}^\xi \rangle$:
this would yield
$\Sigma=\Sigma_{el}+\Sigma_{in}$ with
$\Sigma_{in}=\langle \hat{\Sigma}_{in}^\xi \rangle$, and
the scattering rate $\Gamma=-2 Im \Sigma$ would indeed separate into a sum
of two contributions from the phonons and from disorder respectively, 
$\Gamma= \Gamma_{el}+\Gamma_{in}$.
In general, however, one must use the fully disorder-dependent  scattering rate, that we write here as 
\begin{equation}
\Gamma= \Gamma_{el}+ \Phi \Gamma_{in},
\label{eq:GammaPert}
\end{equation}
having defined the dimensionless vertex function
 $\Phi=\frac{\langle \hat{G}_{el}^\xi \hat{\Sigma}_{in}^\xi \hat{G}_{el}^\xi\rangle}{
\langle  \hat{G}_{el}^\xi \rangle^2  \langle  \hat{\Sigma}_{in}^\xi \rangle}$  
which embodies lattice/disorder correlations.
Explicit deviations from Matthiessen's rule arise as soon as $\Phi<1$. 
It should be stressed that we are implicitly ignoring here additional deviations that can arise  when 
the self-energy has a non-trivial matrix structure (due e.g. to non-local effects, not included in 
Eq. (1), or to multi-band effects beyond our simple band model).

The lowest order inelastic contribution (phonon exchange)
entering in Eq. (\ref{eq:Dyson}) is readily obtained as 
$\hat{\Sigma}_{in}^\xi=s^2 \hat{G}_{el}^\xi$, highlighting the dependence 
of inelastic scattering on the
local environment variables.
The prefactor $s^2$ 
measures the fluctuations of the site energy induced by the  atomic motions. 
These motions are thermal above the Debye temperature, leading to $s^2\propto T$ owing to 
the equipartition principle, which ultimately causes a linear resistivity
$\rho\propto \Gamma_{in}\propto T$  (an explicit calculation 
assuming a local interaction with Einstein phonons is provided in the SM file, p.11).
The above expression for $\hat{\Sigma}_{in}^\xi$, together with Eq. (\ref{eq:GammaPert}),
allows us to explicitly relate the TCR to
the properties of the disordered metal at $T=0$. 
In particular, neglecting the weak temperature dependence of $\Gamma_{el}$ in Eq. (\ref{eq:GammaPert})
we obtain:
\begin{equation}
\frac{d\Gamma}{dT}\simeq  \frac{d\Gamma_{in}}{dT}  \times \Phi.
\label{eq:GammaPert2}
\end{equation}
This  expression is very appealing because it separates the temperature dependence 
of the resistivity into a conventional term $\frac{d\Gamma_{in}}{dT}>0$ connected with
the inelastic scattering off the phonons in the absence of disorder,
and a 
factor $\Phi=  \langle (\hat{G}_{el}^\xi)^3 \rangle/ \langle \hat{G}_{el}^\xi \rangle^3 $ 
controlled by 
disorder alone, 
and  which is  entirely responsible for  the sign of $d\Gamma/dT$.

\begin{figure*}[t!]
\centering
\includegraphics[width=16cm]{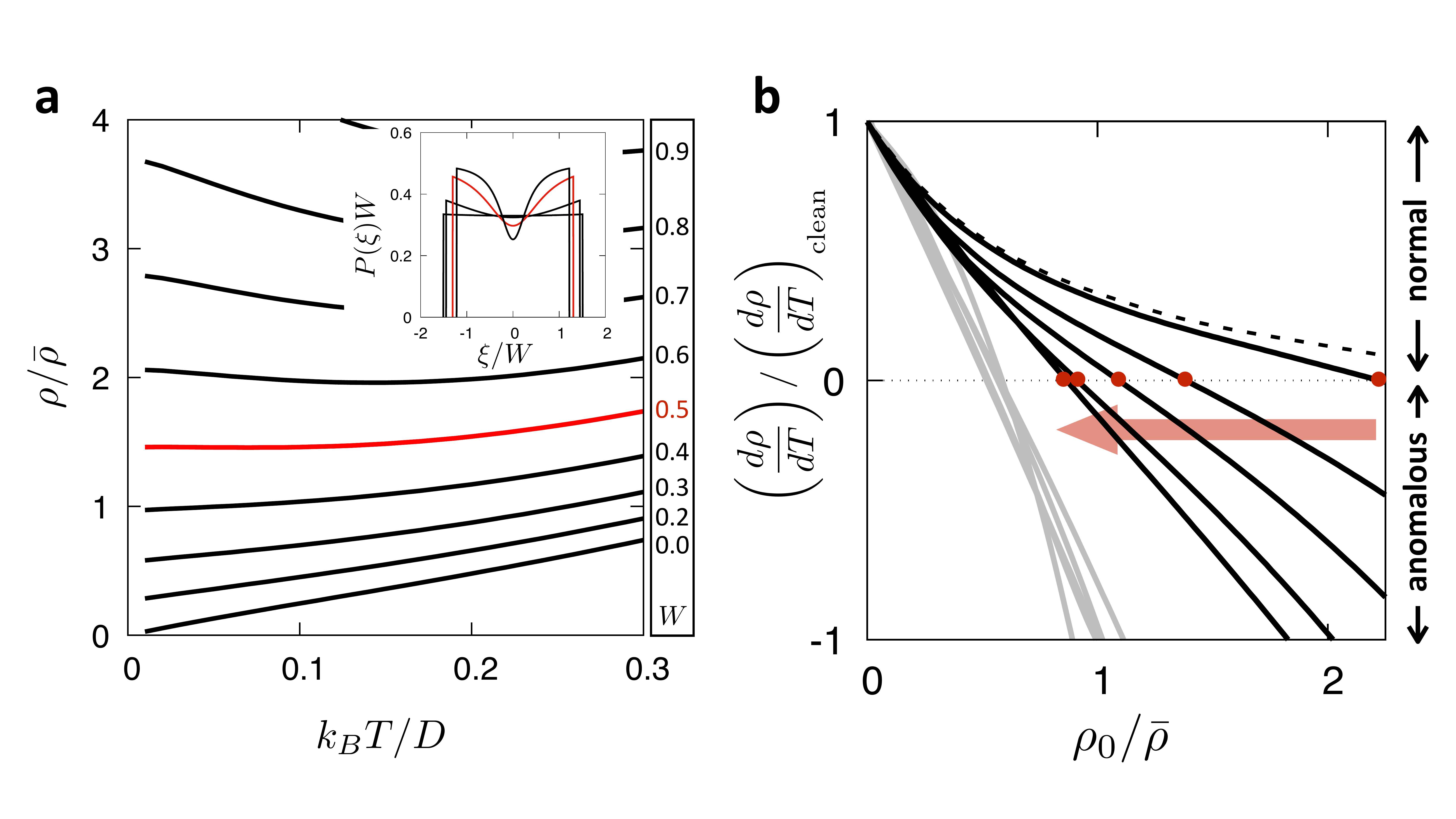}
\caption{\textbf{Numerical results.}
(a) Resistivity vs. temperature in the uniform disorder model at fixed $\lambda=0.2$, for increasing disorder strengths 
$W$ as indicated by the labels
(temperature and $\rho$ units are set respectively by the half-bandwidth $D$ and 
$\bar\rho=a\hbar/e^2$). The flat resistivity curve is marked in red.
Inset: The corresponding renormalized site energy distribution 
$P(\xi)$ at $T=0$, for  selected values of  
$W=0.0$ (bare distribution, see text),  $0.2, 0.5, 0.8$.
(b) The rate of variation $d\rho/dT$  
extracted from the resistivity data in the linear range $k_BT/D=0.01-0.05$
normalized by its value in the clean limit as a function of $\rho_0/\bar\rho$, for the uniform disorder model
(black lines) The different curves 
represent increasing values of $\lambda=0.05,0.2,0.3,0.4,0.5$, in the order indicated by the 
arrow, spanning essentially  the entire metallic region allowed by the theory (see text).
The  dashed line indicates the $\lambda\to 0$ limit. 
Red dots mark the locus $\rho^*$ of flat resistivity. The results for the binary disorder model, shown as gray lines,
do not exhibit appreciable $\lambda$-dependence in the region under study. 
}
\label{fig:rhos}
\end{figure*}

To demonstrate 
the emergence 
of Mooij correlations, 
we now expand Eq. (\ref{eq:GammaPert2}) in the weak disorder limit.
Both $\Phi$ and $\Gamma_{el}$ can be explicitly evaluated
by expressing  the local Green's function 
$\hat{G}_{el}^\xi=(G_0^{-1}-\xi)^{-1}$ in terms of the Weiss field  
$G_0^{-1}$, which embodies the propagation from a given site to the rest of the lattice. 
Having checked numerically that the results do not change qualitatively as the band filling is varied within the metallic phase, 
we present here the particle-hole symmetric case for simplicity. In this case $G_0^{-1}=i\nu$ is purely imaginary, where $\nu$ 
defines the escape rate from an atomic site, 
which leads to $\Phi\simeq 1-3\Gamma_{el}/2\nu$ for $\Gamma_{el}\ll \nu$,
with $\Gamma_{el}=2\langle\xi^2\rangle/\nu$.
The above expansion 
is totally general: when combined with Eq. (\ref{eq:GammaPert2}), it states that 
the temperature dependence of the 
scattering rate in a weakly disordered metal is entirely 
known in terms of 
its behavior in the clean limit and of the residual zero temperature value $\Gamma_{el}$. 
Moreover, the relationship is governed by
a single number $\nu$ which is proportional to the average density of states (DOS) at the Fermi energy
(from the Fermi golden rule, 
$\nu\sim t^2 N(E_F)\sim t^2/t$ with $t$ the inter-atomic transfer integral).   
Substituting into Eq. (\ref{eq:GammaPert2}) and 
using the Drude formula 
to convert to resistivities  
directly implies 
\begin{equation}
\rho(T)=\rho_0 + (\rho^*-\rho_0) AT,
\label{eq:Mooijpert}
\end{equation}
with $A$ and $\rho^*$ material-specific constants. 
In the  weak disorder limit where the above formula applies, $\rho \lesssim \rho^*$, the TCR
is indeed linearly correlated with the residual value $\rho_0$.

The foregoing derivation
leads us to the following important conclusions. First, 
it demonstrates that neither the Mooij correlations nor the
existence of an anomalous metallic regime with $\frac{d\rho}{dT}<0$ require 
to be in the immediate proximity to a metal-insulator transition. 
This is further confirmed by the fact that accounting for 
Anderson localization using the more general method of Ref. \cite{DiSantePRL17}
does not alter the overall picture described here (cf. Fig. S4 in the SM file).
Quite on the opposite,
the Mooij phenomenon is rooted 
in those correlations between inelastic and elastic scattering
that arise already at weak disorder, and
which are also responsible for the  breakdown of Matthiessen's rule. 
Second, in those cases where 
the decrease predicted by Eq. (\ref{eq:Mooijpert})  extends 
all the way down to the point where the slope $\frac{d\rho}{dT}$ changes sign,
then the T-independent resistivity is predicted to 
occur when the scattering rate is of the order of a 
fraction of the escape rate. Since this is also  a fraction of the electronic bandwidth, 
the above argument shows that the condition for flat resistivity and the MIR limit qualitatively coincide,
as was long assumed.

\subsection*{Polaronic renormalization of disorder}
The occurrence of an anomalous metallic regime
where $d\rho/dT$ changes sign
sets precise conditions on the form of the actual disorder distribution $P(\xi)$.
From the general expression of 
$\Phi\propto -\int d\xi \frac{1}{\nu^2+\xi^2} 
\frac{d^2P}{d\xi^2} $ (Eq. (S19) in the SM file),
it is apparent that for sufficiently strong disorder, i.e.
when the fluctuations of the site energies $\xi$ become large on the scale of the escape rate $\nu$, 
the sign of the correlation vertex $\Phi$ is fully determined 
by the curvature of $P(\xi)$ around $\xi=0$.
Given that $ d\rho/dT\propto \Phi$, a sufficient condition for the emergence of negative TCR
is therefore the existence of a dip in the disorder distribution around the Fermi energy. 
Conversely, according to this same argument a
featureless distribution of site energies as is commonly assumed in theoretical studies of 
disordered systems\cite{Anderson}
is  unable to yield such a change of sign, because $\frac{d^2P}{d\xi^2} =0$.
Similarly, no anomalous temperature dependence should arise in the case of
gaussianly distributed site energies ($\frac{d^2P}{d\xi^2} <0$), 
as realized for example in systems hosting randomly 
distributed charges or dipoles \cite{Galitski}.
The dimensionless 
function $\Phi(\Gamma_{el}/\nu)$ for different distributions of disorder 
is illustrated in Fig. S1 of the SM file.

The prediction of  
a large variability of behaviors 
depending on the form of disorder seems at odds 
with the  general experimental observation of resistivities with negative temperature coefficients 
in sufficiently disordered metals. 
The origin of this 
widespread behavior lies 
in the response of the deformable lattice,
which is able to modify the very nature of elastic scattering (Fig. \ref{fig:sketches} (c,d)): 
triggered by the existing randomness, 
polaronic deformations will inevitably arise and 
convert any given, even featureless, disorder distribution into one with universal 
characteristics \cite{DiSantePRL17}.
Furthermore, such polaronic renormalization of the disorder potential 
is not restricted to materials
with particularly strong electron-phonon interactions but is expected
to occur in all metals, 
even those which have nominally weak interactions.
This can be understood as follows. In the presence of a non-vanishing 
electron-phonon coupling, the lattice locally distorts 
in response to the inhomogeneities of the electron liquid. 
This gives rise to an induced potential $\Sigma^H(\epsilon)$ 
which adds to the  random site potential $\epsilon$, so 
that the total potential felt by the electrons is now the sum $\xi=\epsilon +\Sigma^H(\epsilon)$.
The induced potential behaves exactly as the induced magnetization 
in an applied external magnetic field, being $\Sigma^H(\epsilon)\propto \mathrm{sign}(\epsilon)$ at large $\epsilon$ \cite{DiSantePRL17}: it is 
attractive for sites whose energy is below the Fermi level, 
and repulsive above the Fermi level, 
so that it effectively shifts the site energies away from $\epsilon=0$, 
separating the lattice sites into those which host a polaron and those which don't.
As a result, upon inclusion of the response of the deformable lattice,
the actual  distribution $P(\xi)$ of the renormalized local potentials
inevitably develops a dip  around $\xi=0$  (Fig. \ref{fig:sketches} (d)). 
According to the general arguments given in the preceding paragraph,
this eventually enables $d\rho/dT<0$
{\it regardless} of the shape of the initial distribution of $\epsilon$.
We stress that the inclusion of dephasing or inelastic scattering effects 
which do not affect the real part of the self-energy, are instead unable  to modify the effective disorder distribution
\cite{Logan,Girvin}.

\subsection*{Numerical results}

\begin{figure*}[t!]
\centering
\includegraphics[width=16cm]{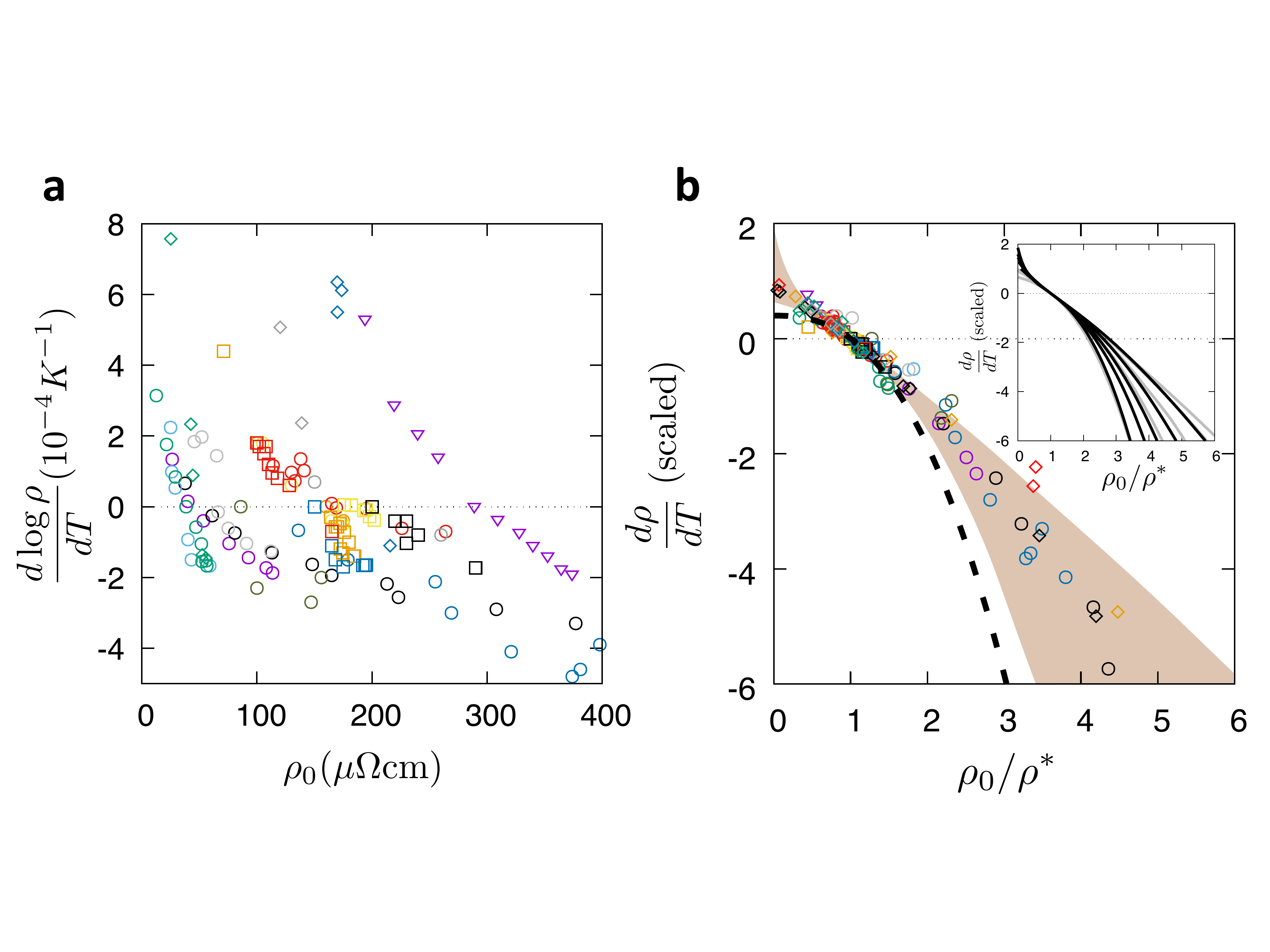}\\
\vspace{-2cm}
\includegraphics[width=12cm]{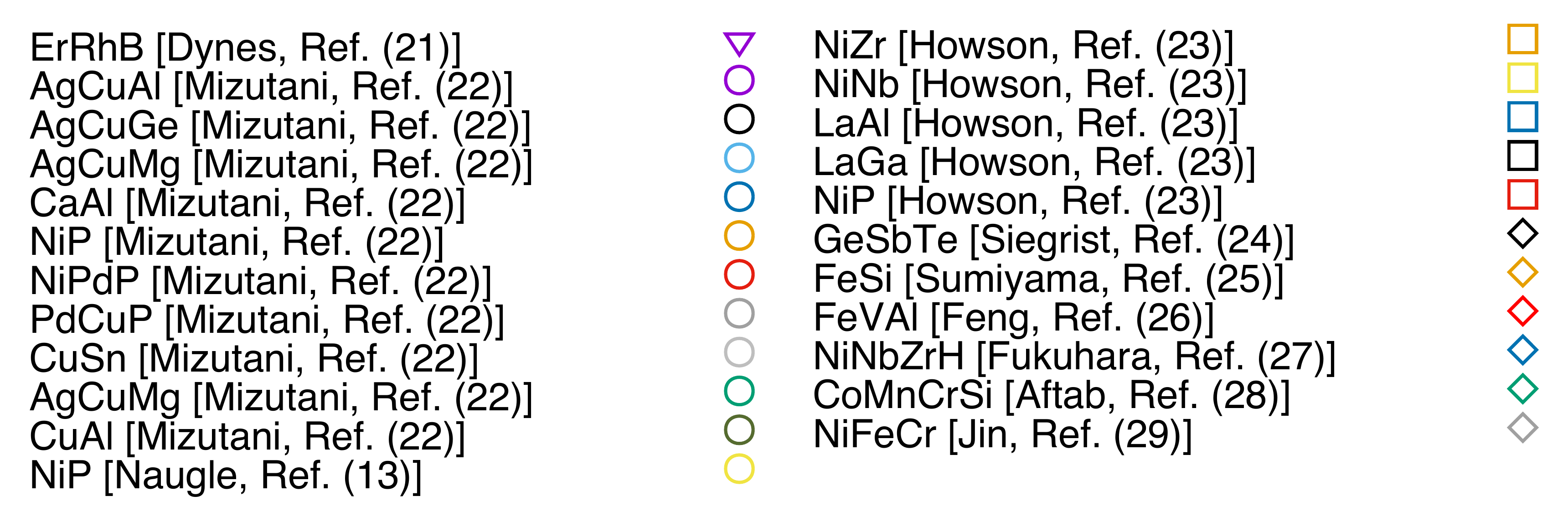}
\caption{\textbf{Comparison with experimental data.} 
(a) A collection of experimental data available in the literature,
including data from Ref. \cite{Tsuei} and more recent references.
We have selected all series of data known to us which display a monotonic decrease
of $d\rho/dT$ and a change of sign upon increasing randomness.
(b) Same data, where each series has been rescaled as suggested in the text, 
allowing for comparison of disordered metals of a different chemical nature in the same plot.
The inset reports the theoretical curves of Fig. \ref{fig:rhos}(b) 
scaled with the same procedure. Their span represents the whole range allowed by 
theory within the metallic phase,  which is reported 
in the main plot as a shaded area. The 
dashed line is the weak-localization theory of Refs. \cite{Kaveh,Tsuei}.
Triangles (from [Dynes], Ref. \cite{Dynes}) and circles
(from [Mizutani], Ref. \cite{Mizutani} and [Naugle], Ref. \cite{Naugle}
correspond to the data already reported in Fig. 1 of Ref. \cite{Tsuei}.
Squares are from [Howson], Ref. \cite{Howson}. The remaining data are from 
[Siegrist], Ref. \cite{Siegrist},
[Sumiyama], Ref. \cite{Sumiyama},
[Feng], Ref. \cite{Feng},
[Fukuhara], Ref. \cite{Fukuhara},
[Aftab], Ref. \cite{Aftab} and
[Jin], Ref. \cite{Jin}. 
As the stoichiometry of the compounds can vary within each series, only the constituent elements are indicated.}
\label{fig:exp}
\end{figure*}

Having illustrated the physical mechanisms at play in full generality, we now proceed by
studying two representative models for disordered metals using 
the fully self-consistent DMFT-CPA method.
Fig. \ref{fig:rhos}(a) reports $\rho(T)$  calculated  considering a 
uniform initial distribution of site energies $P_0(\epsilon)=\theta(W^2-\epsilon^2)/(2W)$ 
for a generic metal characterized by a featureless semi-elliptical DOS of half-width $D$
and an  electron-phonon interaction of moderate strength, $\lambda=0.2$ (we consider here an onsite 
interaction with dispersionless phonons characterized by a force constant $K$ and 
coupling constant $g$, 
so that $\lambda = g^2/(KD)$, see Eq. (S1) in the SM file for details).
We conveniently express the resistivity in units of $\bar\rho=a\hbar/e^2$ ($\sim 10^2 \mu \Omega cm$ 
for typical simple metals, assuming a typical lattice spacing $a\simeq 3\AA$),  which is of the order of the MIR limit.
The resistivity curves  reproduce
the typical phenomenology observed in experiments, both qualitatively and quantitatively \cite{LeeRMP}.
In particular,  
the flat resistivity (red curve) occurs for values close to $\bar\rho$ as anticipated.
The evolution of the TCR as a function of the zero temperature intercept $\rho_0$ 
is reported in Fig. \ref{fig:rhos}(b) (black full lines, different curves corresponding to different values of $\lambda$).
At weak disorder, all curves do tend to a common linear behavior 
as predicted by our analytical derivation.

We now explore the dependence on the electron-phonon interaction strength, spanning the whole metallic regime 
up to the breakdown of the metal, which is delimited at large $\lambda$ by 
the formation of a polaronic insulator (the critical value for the polaronic transition
is $\lambda_P=0.67$ in the clean limit \cite{MillisPRB1996}, which is progressively reduced
by the inclusion of disorder \cite{DiSantePRL17}).
The black dashed curve is the limiting behavior obtained in the uniform disorder model for 
vanishingly weak electron-phonon interactions, $\lambda\to 0$. 
In the absence of electron-phonon interactions 
the flat distribution $P_0(\epsilon)$  does not allow for negative values of $d\rho/dT$, so that 
the initial  decrease at low $\rho_0$ progressively flattens out and 
saturates for strong scattering.
The situation changes radically as soon as 
$\lambda\neq 0$.
As shown in the inset of Fig. \ref{fig:rhos}(a), the distribution $P(\xi)$ is  
progressively depleted at $\xi\simeq 0$
owing to the buildup of polaronic lattice deformations; 
this enables $d\rho/dT<0$ above a finite $\rho^*$, precisely as demonstrated in 
the preceding paragraphs.
We see from Fig. \ref{fig:rhos}(b) that the polaronic effects responsible for the anomalous metallic behavior
are  crucial already for quite modest interaction strengths, typical of metals:  
the locus $\rho^*$ of flat resistivity (red dots), 
which diverges for $\lambda\to 0$, 
rapidly decreases with $\lambda$ and reaches values of the order of
the resistivity unit $\bar\rho$ already for $\lambda=0.05$.
This demonstrates that, aside from specific cases where
the electron-phonon coupling is particularly weak,
the flat resistivity can generally be identified with the MIR limit.
Polaronic effects obviously get stronger with increasing $\lambda$, eventually making the 
disorder distribution  markedly bimodal\cite{MillisPRB1996}. 
Correspondingly, 
the calculated behavior gradually approaches that obtained  from a binary distribution 
of site energies, as realized for example in binary alloys 
($P_0^{BM}(\epsilon)=[\delta(\epsilon-W) +\delta(\epsilon+W)]/2$, gray lines).
We note that in the case of binary disorder,
the data in the explored range 
show little dependence on the electron-phonon interactions: 
the distribution of site energies is already bimodal in the absence of interactions, 
which is not altered by the presence of the deformable lattice.

\section*{Discussion}

We finally address the question of weak localization corrections, proving --- based on 
an accurate comparison with existing experiments --- 
our initial assertion that these are not the main cause of the observed
of Mooij correlations.
Mott and Kaveh\cite{Kaveh} were the first to suggest that weak localization physics could lead to a
change of sign of the TCR. Their results were subsequently used by Tsuei \cite{Tsuei}
to argue that the behavior of  TCR across different materials is non-universal, contrary to the
initial claims by Mooij\cite{Mooij}. A more careful analysis, however, shows that 
the formulation of Ref. \cite{Kaveh} does indeed predict
a universal TCR behavior, provided that one properly takes out all material-specific properties 
(a well-known analogy is that of phase transitions, which exhibit universal behavior only after the temperature
is properly rescaled with the material-specific critical temperature $T_c$). In particular,  
since the value of the "flat" resistivity $\rho*$ varies from material to material, 
a meaningful scaling procedure --- both for  theory, and for the analysis of experimental data --- 
should start by expressing resistivities in units of  $\rho*$.
Following this idea, we were able to show that 
the weak localization formulas of Refs. \cite{Kaveh,Tsuei} assume the following 
universal form (details in Sec. 2 of the
SM file):
\begin{equation}
\mathrm{TCR}(\rho_0/\rho^*)= \frac{2C}{5} \left[1-(\rho_0/\rho^*)^{5/2}\right] \; \; \; \mathrm{(weak \ loc.)}
\label{eq:TCRsca}
\end{equation}
where $C$ is a  prefactor
which sets the slope of the curve TCR$(\rho_0/\rho^*)$ at $\rho_0=\rho^*$, and which also depends on 
material-specific properties (e.g. the strength of the electron-phonon interactions and the carrier density).

Independently of the validity of Eq. (\ref{eq:TCRsca}),
the scaling arguments provided above  
point to a better way of collecting the transport characteristics of different materials than  
reporting $\frac{d\log\rho}{dT}$ as a function of the zero-temperature intercept
 $\rho_0$ as is customarily done\cite{Mooij,Tsuei} 
(see Fig. \ref{fig:exp}(a)), 
by highlighting instead precisely those features of the TCR which are material independent.
This is done in Fig. \ref{fig:exp}(b), where we replot the experimental data for different compounds 
by rescaling the resistivities to the 
corresponding $\rho^*$, and then normalizing the slope to a common unit value.
The first observation is that, in contrast with the original "Mooij plots",
the data now show an almost complete collapse
indicating that the proposed scaling idea is very powerful in rationalizing the experimental data of the whole
class of disordered metals.
The second observation is that
the experiments markedly disagree with the weak localization predictions, both quantitatively and qualitatively: 
not only do the data deviate appreciably from the behavior given by
Eq. (\ref{eq:TCRsca}), but they also show significant scatter, 
(i.e., significant non-universality), incompatible with the weak localization scenario. 
In Fig. \ref{fig:exp}(b) we also report our theoretical results based on the mechanism of polaronic 
disorder renormalization proposed in the present work, by applying an analogous scaling procedure to 
the data of Fig. \ref{fig:rhos}(b): individual curves obtained for different values of the 
electron-phonon interaction strength and different models of disorder are reported in the inset; 
the full extent of the region allowed by theory is identified by the area spanned by all calculated curves, 
and is reported in the main panel
(shaded area). As opposed to the weak localization scenario,  in this case
both the locus and scatter of data points are in excellent agreement with the data.
Unscaled fits to the original experimental data set of Ref. \cite{Tsuei} 
are shown in Fig. S3 of the SM file, illustrating 
the much better level of agreement achieved in that case by the present model as compared to weak localization theory.

The polaronic mechanism  of strong disorder renormalization
identified in this paper appears to provide a general and 
quantitatively accurate explanation for the mysterious high-temperature transport anomalies \cite{LeeRMP}, 
which are commonly observed in strongly disordered metals \cite{Tsuei}. It is interesting to note that 
a very similar method of expanding around the MIR limit,  again performed within the generalized DMFT setup, 
recently shed light \cite{Vucicevic2015}  on another long-standing puzzle --- that of the high-temperature "Bad Metal" 
behavior in doped Mott insulators. In that situation,  which reflects strong electron-electron 
correlations in absence 
of disorder, one again finds linear-T resistivity, but its slope remains positive on both 
sides of the MIR limit. 
What happens when disorder
and strong electron-electron interactions 
coexist remains a fascinating open question we reserve for future work.
At any rate, we speculate that the  mechanism illustrated here
could very generally apply, leading to anomalous resistivity behavior 
whenever a "deformable" medium is present --- atomic, electronic, or magnetic --- 
that is able to locally respond to disorder.

\section*{Data availability}
The data sets generated during and/or analysed during the current study are available from the corresponding author on 
reasonable request.


\section*{Acknowledgments} V.D. was supported by the NSF grant DMR-1410132. 
D. Di S. acknowledges the German Research Foundation (DFG-SFB 1170), 
the ERC-StG-336012-Thomale-TOPOLECTRICS and
the Gauss Centre for Supercomputing e.V. (www.gauss-centre.eu) for funding
this project by providing computing time on the GCS
Supercomputer SuperMUC at Leibniz Supercomputing
Centre (LRZ, www.lrz.de)

\section*{Competing interests}
The authors declare no competing interests.

\section*{Author contributions}

S.C., V.D. and S.F. led the project and co-wrote the paper. D.D.S. and S.C. implemented the method. 

\end{document}